# A Measure of Flow Vorticity with Helical Beams of Light

Aniceto Belmonte*, [1], Carmelo Rosales-Guzmán[2], and Juan P. Torres[1, 2]

[1]Technical University of Catalonia, BarcelonaTech, Dept. of Signal Theory and Communications
08034 Barcelona, Spain

[2]ICFO–Institut de Ciencies Fotoniques, Mediterranean Technology Park
08860 Castelldefels (Barcelona), Spain

*belmonte@tsc.upc.edu

Vorticity describes the spinning motion of a fluid, i.e., the tendency to rotate, at every point in a flow. The interest in performing accurate and localized measurements of vorticity reflects the fact that many of the quantities that characterize the dynamics of fluids are intimately bound together in the vorticity field, being an efficient descriptor of the velocity statistics in many flow regimes. It describes the coherent structures and vortex interactions that are at the leading edge of laminar, transitional, and turbulent flows in nature [1]. The measurement of vorticity is of paramount importance in many research fields as diverse as biology microfluidics, complex motions in the oceanic and atmospheric boundary layers, and wake turbulence on fluid aerodynamics. However, the precise measurement of flow vorticity is difficult [2]. Here we put forward an optical sensing technique to obtain a direct measurement of vorticity in fluids using Laguerre-Gauss (LG) beams, optical beams which show an azimuthal phase variation that is the origin of its characteristic non-zero orbital angular momentum [6]. The key point is to make use of the transversal Doppler effect [3-5] of the returned signal that depends only on the azimuthal component of the flow velocity along the ring-shaped observation beam. We found from a detailed analysis of the experimental method that probing the fluid with LG beams is an effective and simple sensing technique capable to produce accurate estimates of flow vorticity.

Vorticity is defined as the curl $\vec{\omega} = \vec{\nabla} \times \vec{U}$ of a velocity vector field $\vec{U}$. It is a measure of the amount of angular rotation of a material point about a particular position in a flow field, and it may be regarded as a measure of the local angular velocity of the fluid [1]. Typically, vorticity measurement systems are designed to probe one of the components of the vorticity vector $\vec{\omega}$. For simplicity, and without loss of generality, we consider a two-dimensional flow in the $xy$–plane, where $\vec{U} = (U, V, 0)$ and the vorticity is $\vec{\omega} = (0, 0, \omega)$. The z–axis component is defined on the basis of velocity derivatives as $\omega = \partial V / \partial x - \partial U / \partial y$.

Several optical methods for measuring vorticity in a flow – Particle Image Velocimetry (PIV) [6-8] and Laser Doppler Velocimetry (LDV) [9, 10], among others– attend to determine first the instantaneous velocity components $U$ and $V$, and then differentiate velocity data to yield the vorticity field component $\omega$. By whatever means the velocity measurements are estimate, they must be made simultaneously over several closed space locations from which spatial gradients $\partial V / \partial x$ and $\partial U / \partial y$ can be evaluated using finite difference schemes. As any real measure needs to consider the length scales $dx$ and $dy$, flow measurements are always integrated over some area and a spatial average vorticity is measured. The accuracy of the calculated vorticity depends on the spatial resolution $dx$ and $dy$ of the velocity sampling and the uncertainty error on estimates of velocity differences $dU$ and $dV$.

To overcome the shortcomings of finite difference methods for vorticity measurements, other measurement methods use imaging techniques to observe local rotation of the flow. Vorticity Optical Probing (VOP) uses Gaussian laser beams to illuminate the passage of probe particles embedded in the flow to obtain, by image analysis, information about their trajectories [12]. Although small probe particles suspended in a flow will react to fluctuations of rotation in the flow, allowing the vorticity of the flow to be probed directly as it moves along a streamline, it is not always possible –or even convenient– to implant probe particles into the fluid whose dynamics needs to be characterized.

Here we consider a new optical sensing technique that is not dependent upon direct velocity measurements or the use of probe particles, and directly measures the local vorticity of fluid elements. The technique we propose here is akin to laser Doppler anemometry using LG laser beams [6] to illuminate the flow and obtain, by observing the transversal Doppler effect in the reflected signal [3-5], information about vorticity (see Fig. 1).

The vorticity is closely related to the flow circulation. The circulation about a closed contour in a fluid is a scalar integral quantity and measures the macroscopic rotation for a finite area of the fluid. It is defined as the line integral $C_o = \oint \vec{U} \cdot d\vec{l}$ evaluated along the contour of the component of the velocity vector $\vec{U} = (U, V)$ that is locally tangent to the contour. If circulation is considered along a closed circular loop of radius $\rho_0$ lying in the $xy$–plane, it results $C_o = \int_0^{2\pi} U_\phi(\rho_0, \phi) \rho_0 \, d\phi$, where we make use of cylindrical coordinates $\vec{\rho} = (\rho_0, \phi)$ along the circular contour and $U_\phi = -U\sin\phi + V\cos\phi$ represents the azimuthal component of velocity of the fluid. This component of velocity defines the local angular velocity of the flow $\Omega = U_\phi(\rho_0, \phi)/\rho_0$. Interestingly, for a specific circular flow area, the Stoke's theorem $C_o = \iint_A (\vec{\nabla} \times \vec{U}) \cdot \hat{n} \, dA$, where $\hat{n} \equiv \hat{z}$ is the unit vector normal to the loop, allows to express the average normal component $\omega = (\vec{\nabla} \times \vec{U}) \cdot \hat{n}$ of the vorticity field as the circulation $C_o$ of the velocity field around the considered fluid element divided by the loop elemental area $\pi\rho_0^2$. It results a simple expression for the flow vorticity in terms of a line integral of the azimuthal components $U_\phi$ of velocity.

$$\omega = 1/\pi\rho_0^2 \int_0^{2\pi} U_\phi(\rho_0, \phi) \rho_0 \, d\phi. \tag{1}$$

We attempt to measure the flow integral in Eq. (1) by optical means using LG light beams.

Let us assume that a paraxial LG light beam propagating along the $z$ axis illuminates a system of non-interactive, independent small scatterers moving with the flow with velocity $\vec{U}$ and undergoing translation relative to the scattering volume defined by the illumination beam (see Fig. 1). The incident radiation at the transverse position $\vec{\rho}$ of scatters across the beam wavefront can be written as

$$E_{\perp}(\vec{\rho}, t) = E_0(\rho) \, exp\{-i \, [2\pi f t - \Phi(\vec{\rho})]\} \,. \tag{2}$$

For an incident LG laser beam with radial mode number $p = 0$, arbitrary azimuthal mode number $m > 0$, and beam radius $\omega_0$, the phase $\Phi(\vec{\rho})$ in the transverse profile depends only on the azimuthal angle as $m\phi$ and the intensity distribution $|E_0(\rho)|^2$ describes a central dark spot surrounded by a very narrow, bright ring whose radius of maximum intensity is $\rho_0 = \omega_0 \sqrt{m/2}$ [6]. A moving scatter going through the light ring will observe the azimuthal phase gradient $\vec{\nabla}_{\perp}\Phi = m \, \hat{\phi}/\rho_0$ defined by the LG beam. Consequently, and due to the transverse velocity $\vec{U}$ of the scatter, the time rate $\vec{\nabla}_{\perp}\Phi \cdot \vec{U}$ of the echo phase signal yields a frequency shift $f_{\perp}$ that is written as [3]

$$f_{\perp}(\rho_0, \phi) = m/2\pi \, U_{\phi}(\rho_0, \phi)/\rho_0 \,. \tag{3}$$

The frequency Doppler shift given by Eq. (3) depends only on the local angular velocity of the scatter in the flow $\Omega = U_{\phi}(\rho_0, \phi)/\rho_0$ along the doughnut shaped observation region $\rho = \rho_0$ defined by the light beam. Equation (3) allows expressing the circulation contour integral of the velocity $U_{\phi}(\rho_0, \phi)$ and the corresponding average vorticity $\omega$ in Eq. (1) in terms of the frequency transversal Doppler shift $f_{\perp}(\rho_0, \phi)$ as

$$\omega = 2/m \int_0^{2\pi} f_{\perp}(\rho_0, \phi) \, d\phi \,. \tag{4}$$

The line integral in Eq. (4) describes the frequency centroid $\langle f_{\perp} \rangle$, the arithmetic mean or average of $f_{\perp}(\rho_0, \phi)$ along the ring–like observation region,

$$\langle f_{\perp} \rangle \equiv 1/2\pi \int_0^{2\pi} f_{\perp}(\rho_0, \phi) \, d\phi \,, \tag{5}$$

and Eq. (4) becomes

$$\omega = 4\pi/m \, \langle f_{\perp} \rangle \tag{6}$$

The vorticity measurement technique we propose here is based on Eq. (6). It determines the vorticity $\omega$ directly from the estimation of the transversal Doppler frequency centroid $\langle f_{\perp} \rangle$ of the signal backscattered by the flow when illuminated by a LG beam with mode number $m$.

The frequency centroid $\langle f_{\perp} \rangle$ estimation is typically based on the spectrum of the observed signal. The return backscattered signal in time –a compound of signals with different frequency Doppler shift triggered by the multiple components of velocity $U_{\phi}(\rho_0, \phi)$ along the annular illumination beam– can be Fourier transformed to define its frequency spectrum. The characteristic return Doppler spectrum is a histogram of Doppler frequency components describing the spectral content of the returned signal and it can be used to calculate the frequency centroid $\langle f_\perp \rangle$ as the average of the frequencies present in the signal.

We use numerical simulations and experiments with selected engineered flows to demonstrate the viability of the proposed method. When a set of independent scatters, moving with velocity $\vec{U}$, passes the ring–like observation region given by Eq. (2), it generates a burst of optical echoes that contributes to the received optical signal. We apply a superposition model for the scattering process that directly gives the complex amplitude of the return signal as the sum of the fields scattered by all the scatters illuminated by the LG beam (see Supplementary Information section for details). The use of a realistic signal model illustrates the dependence of the results on the different experimental parameters and allows addressing the problem of vorticity estimation under the supposition of both additive (receiver) and multiplicative (speckle) noises, those producing great return signal variability. We assume that the Doppler measurement system uses heterodyne detection –the most straightforward to set up experimentally– where the scattered light is coherently mixed on the receiver with a more intense reference beam, which acts as a local oscillator [13].

In order to proceed with the numerical experiments, we simulate the signal returns by direct implementation of the superposition model (Eq. (S1) on the Supplementary Information section). Figure 2 shows the result of our numerical experiments on two different flow patterns. The technique is tested in a steady laminar flow (Fig. 2(a)), in which the flow vorticity is known, and in a complex flow around a circular cylinder immersed in a uniform flow (Fig. 2(b)). The use of realistic numerical experiments illustrates the dependence and the effects of several flow and illumination parameters on the performance of the probing technique. It allows choosing the best measurement parameters and addressing the optimization problem of vorticity estimation. In these experiments, we consider an incident LG laser beam with radial mode number $p = 0$, azimuthal mode number $m = 10$, and beam radius $\omega_0 = 45$ µm. The illuminating beam phase changes from zero to $2\pi$ ten times around the azimuth and the intensity distribution shows a bright ring of radius $\rho_0 = 100$ µm.

Figure 2(a) shows the measurement of vorticity in a laminar pipe flow. In the numerical experiment, the fluid is flowing along the longitudinal y–axis through a closed channel of radius $R = 2$ mm. The transversal velocity $\vec{U} = (0, V)$ of the flow describes a parabolic profile of velocities along the transversal x–axis that varies from zero at the channel ends to a maximum of $V_0 = 4$ mm/s along the center of the channel. The parabolic profile of velocities $V = V_0[1 - (x/R)^2]$ gives the linear vorticity profile $\omega = 2\,V_0/R^2\,x$. The measurements with LG beams reproduce very closely these expected vorticity values.

In a different numerical experiment, Fig. 2(b) shows vorticity in a complex flow created by the unsteady separation of fluid around a cylindrical object located up stream (don't show in the graph). We estimate the velocity field $\vec{U}$ using a numerical tool for flow simulation. From the numerical velocity field $\vec{U} = (U, V)$ we calculate the expected z–component of vorticity as $\omega = \partial V/\partial x - \partial U/\partial y$. The flows on opposite sides of the cylindrical object interact in an extended region and produce a regular circulation pattern. The energy of the vortices is ultimately expended by viscosity as they move further down stream and the regular pattern disappears. The velocity field is pictured in the right graph with a set of streamlines that are tangent to the flow velocity vector. The color scale in the same graph gives an idea of the vorticity magnitude. The left plot compares a measure of flow vorticity with LG beams and the corresponding theoretical expectations. In the simulation, the measurement is realized across the flow, down stream from the cylindrical object.

The feasibility of the proposed method to measure flow vorticity is also verified through the experiments (see Fig. 3). A heterodyne receiver based on a modified Mach-Zehnder interferometer was used for experiments as shown in Fig. 3(a). Using the insight realized by numerical experiments into the problem of vorticity estimation, the operation parameters of the test system were established as described in the Supplementary Information section. In order to emulate different types of flows, we use a Digital Micromirror Device (DMD). A DMD is an array of individually controlled micromirrors that can be switched on and off to define specific spatial and temporal reflection patterns. By controlling which specific mirrors are in the on or off states, and the timing between these states, we can emulate different types of physical trajectories and velocities of reflecting particles moving with a flow. At each position where the particle would be located, light is reflected back to the detector, while no light is reflected elsewhere. This is equivalent to having a two-dimensional flow in a transverse plane. This system is very convenient to demonstrate in the lab the feasibility of the scheme put forward here. It allows emulating different types of flows with good control of the experimentally relevant parameters such as the velocity profile (a supplementary movie shows one of the flows implemented in the DMD).

In flows over stationary flat plates, there is a gradient of velocity as the fluid moves away from the plate, and the fluid tends to move in layers with successively higher speed. In Figs. 3(b) and 3(c), we test the DMD-based experimental setup with two bi-dimensional laminar boundary layer flows characterized by parabolic and linear velocity profiles, respectively. In the experiments, the fluid is flowing along the longitudinal y–axis and the transversal velocity $\vec{U} = (0, V)$ has a maximum of $V_0 \approx 25$ mm/s at a distance $R \approx 6$ mm from the stationary layer. As a parabolic profile of velocities gives a linear vorticity profile, a linear profile $V = V_0 \, (x/R)$ gives a constant vorticity profile $\omega = V_0/R$.

Experimental measurements show the expected linear vorticity profile over the parabolic profile of velocities (Fig. 3(b)) and a constant vorticity profile over a linear velocity profile (Fig. 3(c)). In both cases there are small differences between theoretical and experimental, as all measurements are subject to some uncertainty due to the limited accuracy of the flow definition in the DMD and the concurrent limitations to dynamic speckle reduction. But, in terms of the slow velocity and fast velocity zones, the trends of vorticity rise in parabolic profiles and constant vorticity in linear profiles were almost the same. Both experiments show that the vorticity profiles extracted from the measurements using the least squares approach in a regression analysis (blue, solid lines) are well into the uncertainty limits to the theoretical expectations as defined by the DMD accuracy (red, dashed lines).

In conclusion, the problem of measuring vorticity in a flow has been confronted. We propose an optical technique that uses LG beams, characterized by ring–like intensity distributions and azimuthal phase variations, to sense rotation at every point in a flow. We develop the theoretical background behind the modeling of optical measurement of vorticity in a flow, identifying the required assumptions and input beam parameters. The spectral properties of the return signal, and the spectrum centroid integral in particular, are fundamental to interpretation of experiments used in flow vorticity monitoring. By using numerical simulations and lab experiments, we assess the feasibility of the sensing technique and identify the accuracy of vorticity measurements from return signals affected by target speckle and receiver noise.

REFERENCES

1. G. K. Batchelor, *An Introduction to Fluid Dynamics* (Cambridge University Press, 2000).
2. J M Wallace, and J F Foss, "The Measurement of Vorticity in Turbulent Flows," Annu. Rev. Fluid Mech. **27**, 469-514 (1995).
3. A. Belmonte and J. P. Torres, "Optical Doppler shift with structured light," Opt. Lett. **36**, 4437 (2011).
4. M. P. J. Lavery, F. C. Speirits, S. M. Barnet and M. J. Padgett, "Detection of a spinning object using light's orbital angular momentum," Science **341**, 537 (2013).
5. C. Rosales-Guzmán, N. Hermosa, A. Belmonte, and J. P. Torres, "Experimental detection of transverse particle movement with structured light," Sci. Rep. **36**, 2815 (2013).
6. R. Loudon, "Theory of the forces exerted by Laguerre-Gaussian light beams on dielectrics," Phys. Rev. A **68**, 013806 (2003).
7. R. J. Adrian, "Particle-imaging techniques for experimental fluid mechanics," Annu. Rev. Fluid Mech. **23**, 261–304 (1991).
8. R. J. Adrian, J. Westerweel, *Particle Image Velocimetry* (Cambridge University Press, 2011).
9. S. Yao, P. Tong, and B. J. Ackerson, "Proposal and testing for a fiber-optic-based measurement of flow vorticity," Appl. Opt. **40**, 4022-4027 (2001).

10. Y. Yeh, H. Z. Cummins, "Localized Fluid Flow Measurements with an He-Ne Laser Spectrometer," Appl Phys. Lett. **4**, 176-178 (1964).

11. F. Durst, A. Melling, and J. Whitelaw, *Principles and Practices of Laser- Doppler Anemometry* (Academic Press, 1981).

12. M. Frish and W. Webb, "Direct Measurement of Vorticity by Optical Probe," *J. Fluid Mech.* **107**, 173-200 (1981).

13. T. Fujii and T. Fukuchi (Eds.), *Laser Remote Sensing* (CRC Press, 2005).

ACKNOWLEDGEMENTS

This research was partially funded by the Spanish Department of Science and Innovation MICINN Grant No. TEC 2012-34799. JPT acknowledges support from ICREA (Generalitat de Catalunya) and the program Severo Ochoa from The Government of Spain. C.R.G. would like to thank V. Rodríguez-Fajardo for useful help to emulate the experimental velocity profiles.

AUTHOR CONTRIBUTIONS

A.B. devised the theory, performed the simulations and analyzed data. C.R.G. constructed the experimental optical system and collected data. J.P.T. worked on the theory. All authors participated in the design of the experiment, prepared the figures and contributed to writing the manuscript.

COMPETING FINANCIAL INTERESTS STATEMENT

The authors declare no competing financial interests.

FIGURE LEGENDS

**Fig. 1.** Measure of vorticity in a flow. (a) Here we show the schematic of an experiment in which the local vorticity of a flow can be estimated by probing the fluid with Laguerre-Gauss beams. The proposed measurement technique considers an incident LG laser beam whose phase depends only on the azimuthal angle and the intensity distribution describes a bright ring of light over the flow. When a set of independent scatters, moving with the flow, passes the ring–like observation region, it generates a burst of optical echoes (scattered glow) that contributes to the received optical signal. (b) The inset depicts the illumination beam on the center of the flow channel with a spatially varying phase-gradient (as indicated by color scales). The key point is to make use of the transversal Doppler effect of the returned signal that depends only on the azimuthal component $U_\phi$ of the flow velocity $\vec{U}$ along the ring-shaped observation beam.

We show that the centroid of the transversal Doppler spectrum allows a direct estimation of the flow vorticity over the area illuminated by the light beam.

**Fig. 2.** Numerical experiments on the measurement of flow vorticity. (a) The parabolic profile of velocities (red line, right axis) in a laminar flow gives a linear vorticity profile (blue line, left axis). The measurements with LG beams (triangular markers) reproduce very closely the expected vorticity values. As an illustrative example, we present (inset, right) the frequency signal spectra corresponding to the measurements S1 and S2 in the plot. (b) Vorticity in a complex flow created by the unsteady separation of fluid around a cylindrical object located up stream (don't show in the graph). The flow is pictured with a set of streamlines (white curves) that are tangent to the flow velocity vector. Left: A measure of flow vorticity with LG beams (triangular markers) and the corresponding theoretical expectations (solid line). In the simulation, the measurement is realized across the flow (transversal dashed line on the right graph), down stream from the cylindrical object.

**Fig. 3.** Lab experiments on the measurement of flow vorticity. (a) In the experimental setup, a collimated Gaussian beam is divided by a Polarized Beam Splitter ($PBS_1$) into a reference beam (red line) and a probe beam (green line). The probe beam acquires the desired phase profile after impinging onto a computer-controlled SLM. This structured light (green line) is filtered and made to shine onto a Digital Mirror Device (DMD). The DMD is controlled with a PC to generate on-demand particle flows with different velocity profiles. Light reflected by the particles (blue line) is made to interfere with the reference beam using a beam splitter (BS). The interference signal is captured using two photodetectors ($PD_1$ and $PD_2$) connected to an Oscilloscope. A phase shifter is used to shift our detected signal to 1 KHz. (b) Over a laminar boundary layer flow characterized by a parabolic profile of velocities, a least squares approach in a regression analysis of the measurements (triangular markers) produce a linear vorticity profile (blue line). (c) As before, but now the laminar boundary layer flow is characterized by a linear profile of velocities.

FIGURES

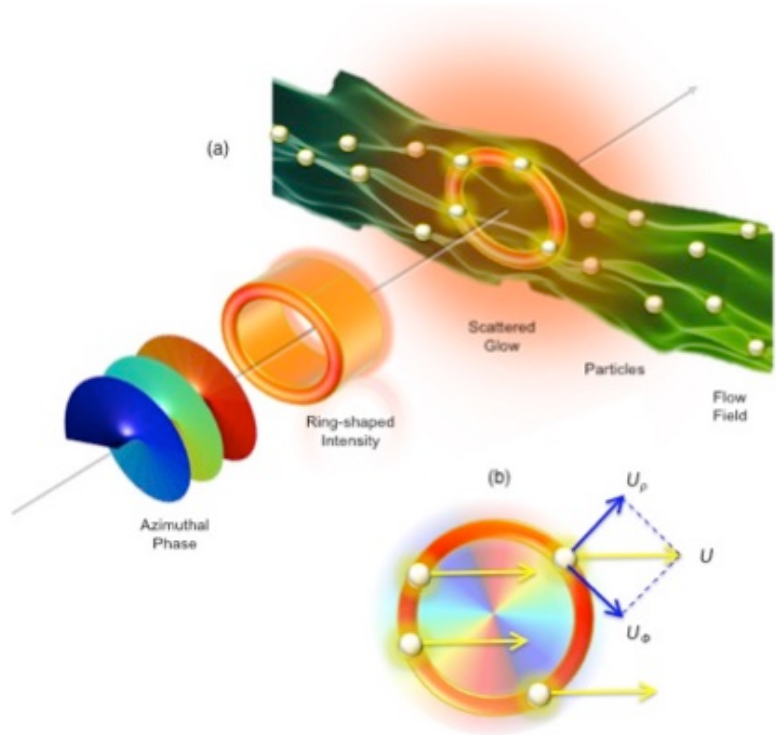

Fig. 1. Measure of vorticity in a flow. (a) Here we show the schematic of an experiment in which the local vorticity of a flow can be estimated by probing the fluid with Laguerre-Gauss beams. The proposed measurement technique considers an incident LG laser beam whose phase depends only on the azimuthal angle and the intensity distribution describes a bright ring of light over the flow. When a set of independent scatters, moving with the flow, passes the ring–like observation region, it generates a burst of optical echoes (scattered glow) that contributes to the received optical signal. (b) The inset depicts the illumination beam on the center of the flow channel with a spatially varying phase-gradient (as indicated by color scales). The key point is to make use of the transversal Doppler effect of the returned signal that depends only on the azimuthal component $U_\phi$ of the flow velocity $\vec{U}$ along the ring-shaped observation beam. We show that the centroid integral of the transversal Doppler spectrum allows a direct estimation of the flow vorticity over the area illuminated by the light beam.

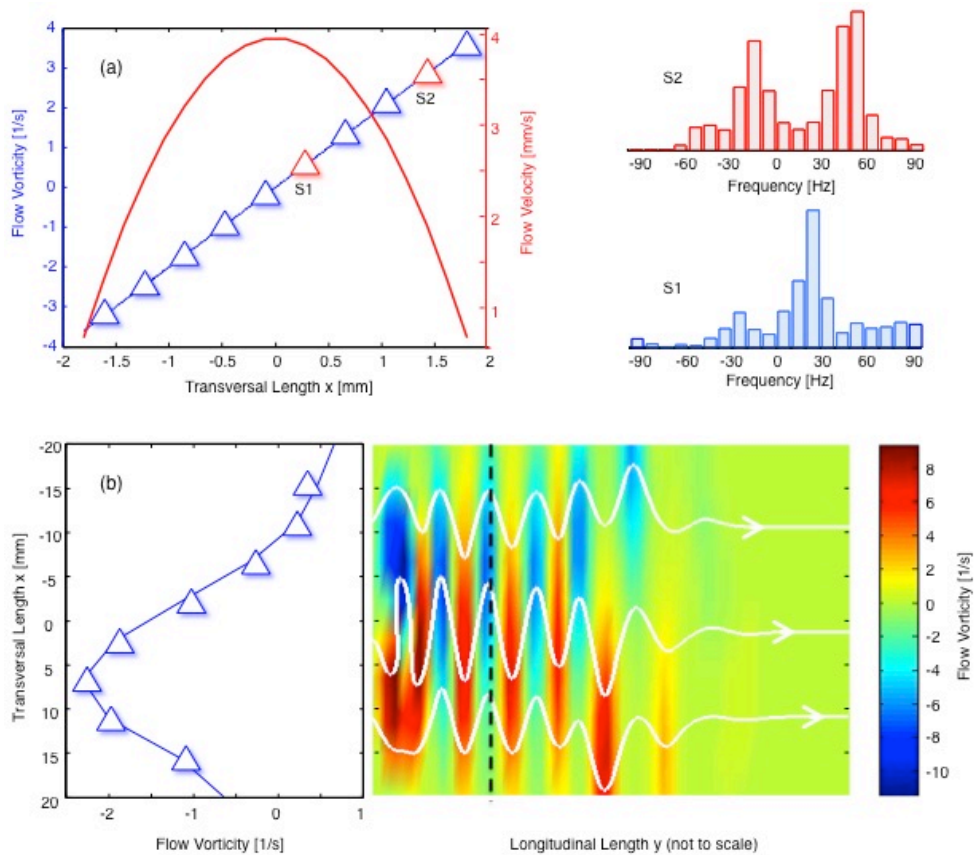

Fig. 2. Numerical experiments on the measurement of flow vorticity. (a) The parabolic profile of velocities (red line, right axis) in a laminar flow gives a linear vorticity profile (blue line, left axis). The measurements with LG beams (triangular markers) reproduce very closely the expected vorticity values. As an illustrative example, we present (inset, right) the frequency signal spectra corresponding to the measurements S1 and S2 in the plot. (b) Vorticity in a complex flow created by the unsteady separation of fluid around a cylindrical object located up stream (don't show in the graph). The flow is pictured with a set of streamlines (white curves) that are tangent to the flow velocity vector. Left: A measure of flow vorticity with LG beams (triangular markers) and the corresponding theoretical expectations (solid line). In the simulation, the measurement is realized across the flow (transversal dashed line on the right graph), down stream from the cylindrical object.

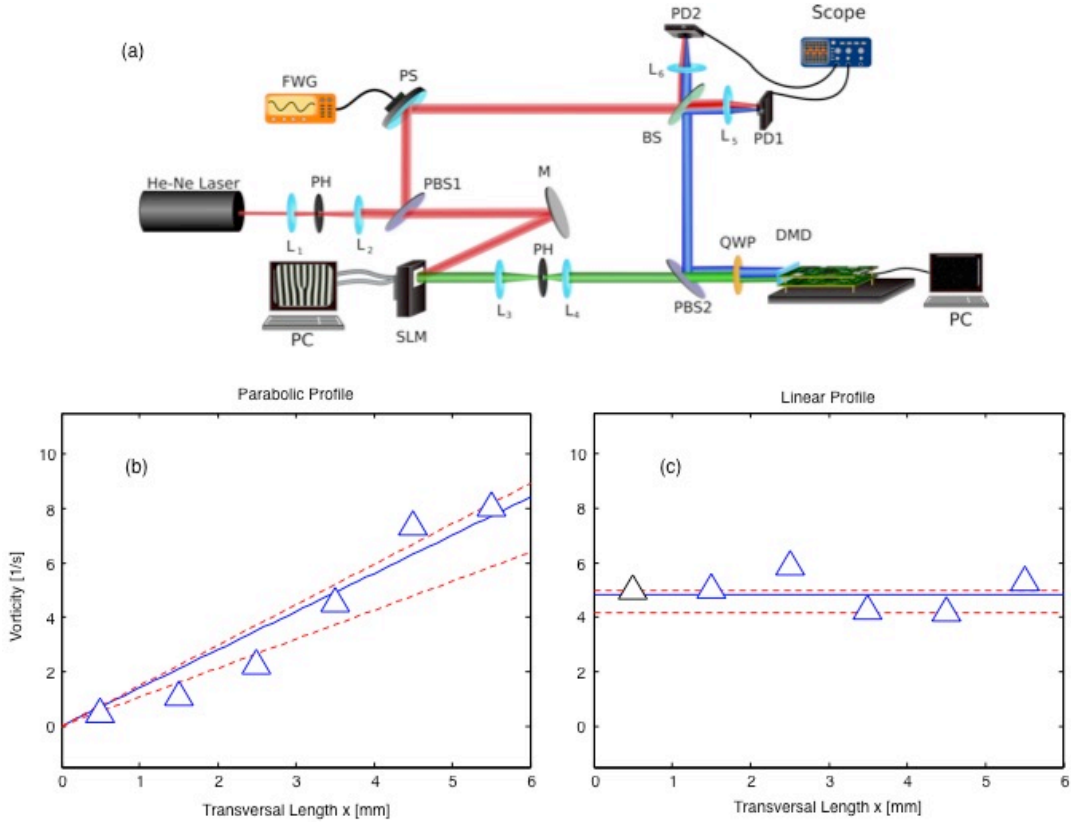

Fig. 3. Lab experiments on the measurement of flow vorticity. (a) In the experimental setup, a collimated Gaussian beam is divided by a Polarized Beam Splitter (PBS$_1$) into a reference beam (red line) and a probe beam (green line). The probe beam acquires the desired phase profile after impinging onto a computer-controlled SLM. This structured light (green line) is filtered and made to shine onto a Digital Mirror Device (DMD). The DMD is controlled with a PC to generate on-demand particle flows with different velocity profiles. Light reflected by the particles (blue line) is made to interfere with the reference beam using a beam splitter (BS). The interference signal is captured using two photodetectors (PD$_1$ and PD$_2$) connected to an Oscilloscope. A phase shifter is used to shift our detected signal to 1 KHz. (b) Over a laminar boundary layer flow characterized by a parabolic profile of velocities, a least squares approach in a regression analysis of the measurements (triangular markers) produce a linear vorticity profile (blue, solid lines). (c) As before, but now the laminar boundary layer flow is characterized by a linear profile of velocities.

SUPPLEMENTARY INFORMATION

## Numerical experiments on the measurement of flow vorticity

When a set of independent scatters, moving with velocity $\vec{U}$, passes the ring–like observation region given by Eq. (2), it generates a burst of optical echoes that contributes to the received optical signal. We use a superposition model for the scattering process that directly gives the complex amplitude of the return signal as the sum of the fields scattered by all the scatters illuminated by the LG beam. After coherent detection and filtering to remove the carrier frequency and its harmonics, we obtain a detected signal (photocurrent) $i$ characterizing the optical echo from the target, which can be written as

$$i = \sum_{j=1}^{n_S} i_{0j}(\vec{\rho}_j) \, exp\{i\,[\psi_j + \Phi(\vec{\rho}_j)]\} + n_0 \,. \tag{S1}$$

Here, the summation is carried out over all the $n_S$ illuminated scatters and $\vec{\rho}_j = (\rho_j, \phi_j)$ is the transverse position of the $j$th scatter. The complex amplitude of the electric signal at the detector $i_{0j}(\vec{\rho}_j)$ takes into account the radiation distribution $E_0(\vec{\rho}_j)$ of the illumination beam, the complex scattering amplitude of the $j$th scatter, and the efficiency of the heterodyne detection process. The phase $\Psi_j = \psi_j + \Phi(\vec{\rho}_j)$ of the return from the $j$th scatter considers two independent terms, with $\Phi(\vec{\rho}_j) = m\phi_j$ describing the scatter position across the beam, and $\psi_j$ the random nature of the return arrival times. As the $n_S$ randomly phased elementary contributions to the total observed field interfere with one another, the resultant intensity is affected by speckle. In Eq. (S1), $n_0$ is an additive noise term being determined by the intensity of the detected signal. In heterodyne receivers, we can consider shot-noise limited signals where the noise power level is proportional to the detected photocurrent $i$ squared.

Equation (S1) above describes not only the instantaneous noisy return, but also the temporal evolution $i(t)$ of the complex amplitude and the dynamics of speckle phenomena. We only need to consider that the instantaneous scatter position vector $\vec{\rho}_j$ changes with the flow velocity $\vec{U}_j$ observed by the $j$th scatter as $\vec{\rho}_j(t) = \vec{U}_j\,t$.

Typically, an heterodyne receiver sample the output signals of a complex receiver with a specific sampling frequency $F_S$ and temporal vector $\vec{\imath} = (i_1, i_2, \ldots, i_M)$ describes the $M$ complex data samples obtained from the return signal $i(t)$ at temporal sampling intervals $1/F_S$. The power Doppler spectrum $I(f_\perp)$ can be estimated from the linear discrete Fourier transform of the complex data vector $\vec{\imath}$ and requires a sufficiently long data sequence for the spectrum to be well defined. The resulting spectral data vector $\vec{I} = (I_1, I_2, \ldots, I_M)$ describes the power levels obtained in $M$ spectral channels $\vec{f}_\perp = (f_{\perp 1}, f_{\perp 2}, \ldots, f_{\perp M})$. For a discretely sampled return, the bandwidth extends over frequencies limited by the Nyquist frequency, which is half the sampling frequency $F_S$. In actual heterodyne receivers aliasing effects are a major concern and signal bandwidth $B$ is a fraction of the system sampling frequency. As the return spectrum is built of frequency components $f_\perp$ arising from sets of scatters moving with different speeds and that are uncorrelated in position, these different frequency components are also uncorrelated.

As speckle noise on the spectral measurements degrades the quality of the spectral data, speckle is usually reduced by summing $N$ independent unsmoothed sample spectrum, and the accumulated spectrum becomes $\vec{I} = \sum_{n=1}^{N} \vec{I}_n$. After spectral accumulation, the signal spectral data $\vec{I}$ can be used to calculate the frequency centroid integral as the weighted mean of the frequencies present in the signal

$$\langle f_\perp \rangle \equiv \int_{-F_S/2}^{F_S/2} I(f_\perp) f_\perp \, df_\perp = \sum_{k=1}^{M} f_{\perp k} I_k \; . \tag{S2}$$

Equation (S2), in conjunction with Eq. (6), allows estimating the vorticity $\omega$ directly from the spectrum of the return signal. In our numerical experiments, typical parameters for a heterodyne receiver are chosen to be signal bandwidth normalized to the sampling frequency $B/F_S = 0.5$, and number of complex samples per estimate $M = 64$. The numerical experiments consider a signal return accumulation (spectral accumulation) of $N = 250$.

## Lab experiments on the measurement of flow vorticity

We extract the Doppler frequency shift imparted by the moving particles via an interferometric technique using the modified Mach-Zehnder interferometer shown in figure 3 (a). A 15-mW continuous wave He-Ne laser (Melles-Griot, $\lambda = 632.8$ nm) is spatially filtered and expanded to a 5 mm diameter beam. This by using a combination of lenses L₁ and L₂ of focal lengths $f_1 = 50$ mm and $f_2 = 200$ mm respectively, and a 30-mm pinhole (PH₁) placed at the middle focus. Afterwards the beam is split into two beams, a reference (red line) and a proof (green line), using a polarizing beam splitter (PBS₁). A mirror (M) redirects the proof beam to a Spatial Light Modulator (SLM) that imprints the beam with the desired phase profile. The first diffracted order of a fork-like hologram encoded into the SLM is used to illuminate a digitally emulated flow, as explained below. The remaining diffracted orders are spatially filtered using two lenses (L₃ and L₄) of focal lengths $f_3 = 150$ mm and $f_4 = 35$ mm respectively and a pinhole (PH₂), 200 mm in diameter, placed at the middle focus. The outer diameter of the resulting Laguerre-Gauss beam is 1160 $\mu m$ when the winding number is $m = 10$. A second polarizing beam splitter (PBS₂) in combination with a quarter waveplate (QWP) collects light reflected from the simulated flow back into the interferometer (blue line). These reflections are afterwards interfered with the reference signal using a beam splitter (BS). A balanced detection scheme is implemented with two photodetectors $PD_1$ and $PD_2$ connected to an oscilloscope (TDS2012 from Tektronix). To eliminate the low-frequency noise, a Phase Shifter (PS) connected to a Frequency generator (FG), placed in the path of the signal beam, shifts our detected frequency to 1 kHz.

We emulated particle flows with different velocity profiles using a DMD (DLP3000 Lightcrafter evaluation mode). The RGB LED light engine was removed to expose the DMD display. The DLP3000 is composed of 415,872 micromirrors (with a diagonal side length of 10.8 $\mu m$) arranged in a diamond pattern geometry 608×684. Each mirror can be tilted individually from $+12°$ to $-12°$. Hence, by carefully aligning the DMD such that $+12°$ coincides with the plane perpendicular to the incident beam, only mirrors tilted $+12°$ will reflect light parallel to the incident beam, defining our "on" state. Mirrors tilted at $-12°$ will reflect light at an angle that is blocked, defining our "off" state. The time in which the micromirrors are in the "on" or "off" state can be controlled by the software provided with the DMD. In order to simulate a particle moving in an specific direction, a set of 1-bit depth images is uploaded in the DMD software, where a "0" corresponds to the "on" state and a "1" to the "off" state. For example, to simulate a particle moving from left to right, we start with an image in which only one mirror close to the left edge is in the "on" state, in subsequent images this "on" state is displaced to the right by turning "off" this mirror and turning "on" the mirror to its right.

In our experiment we simulated different velocity profiles by creating a set of images, so that in the first images approximately 40,000 of the 608×684 available micromirrors are randomly set to an "on" state. In subsequent images these micromirrors are changed to the "off" state and new mirrors to the "on" state. Which mirrors are turned "on" depends on the velocity profile we want to simulate. For example, in the supplementary movie we show a parabolic velocity profile. Notice how the "on" states (in white) close to the edge of the image moves much slower than those at the center of the image. Those particles that leave the image area are replaced by the same amount following the same velocity profile.

### Data analysis

Both the numerical and the lab experiments share the same data analysis. Understanding how the strength of our signal is distributed in the frequency domain is central to the design of any vorticity sensor intended to use the signal Doppler shift. Our data processing and analysis is rather straightforward and computes a spectrum or spectral density starting from a digitized time series, typically measured in Volts at the input of the A/D-converter. We use the overlapped segmented averaging of modified periodograms for power spectrum estimation. In our case, a periodogram is the discrete Fourier transform (DFT) of one segment of the signal time series that has been modified by the application of a time-domain window function. The practical implementation involves equal binning of frequencies, Hanning windowing, filtering of unwanted residual amplitude modulations, and spectrum averaging to reduce the variance of the spectral estimates. Finally, an estimation of the mean Doppler frequency, or Doppler centroid, operates in the power spectrum of the data by applying Eq. (S2).

## Flow Movie

The supplementary movie visualizes one flow generated by the DMD in the experimental setup. The elapsed time is displayed in the top right corner of the movie. Here, x and y are the transversal and longitudinal dimensions of the flow channel, respectively. The illuminating beam intensity distribution is shown as a bright yellow ring of radius $\rho_0 \approx 500$ μm. The parabolic profile of velocities used to describe the flow is exposed in the movie (blue line, right axis).